\documentclass[sort&compress,12pt]{article}%
\usepackage{amssymb}
\usepackage{amsfonts}
\usepackage{amsmath}
\usepackage{breqn}
\usepackage{graphicx}
\usepackage[T1]{fontenc}
\usepackage[utf8]{inputenc}
\usepackage{xcolor}
\usepackage{geometry}
\usepackage{cite}
\begin{document}

\title{Nonrelativistic Spinless Particle in Vicinity of Schwarzschild-like Black Hole}
\author{ R. G. G. Amorim$^{a,b}$, V. C. Rispoli$^{a}$, S. C. Ulhoa$^{b,c}$, K. V. S. Araújo$^{c}$\\ ${}^{a}$ Faculdade UnB Gama, Universidade de Brasília, Brazil\\${}^{b}$Canadian Quantum Research Center,  Canada\\
${}^{c}$ International Center of Physics, Universidade de
Bras\'{\i}lia, Brazil}
\maketitle

\begin{abstract}

In this article we analyze the behavior of a non-relativistic spinless particle near the event horizon of a Schwarzschild-like black hole. In this way, the Schrödinger covariant equation that describes the particle is obtained from the Galilean covariance technique. The Schrödinger equation in a Schwarzschild-like spacetime is solved analytically and its solutions are given in terms of the confluent Heun function. As a relevant result, we discovered that the energy levels of the particles are quantized and that the particle does not escape to infinity. We obtain the existing transmission and reflection coefficients for a particle and anti-particle pair at the event horizon. We thus verify that there is no non-relativistic equivalent of Hawking radiation.


\end{abstract}

\section{Introduction}

The Black holes were firstly speculated by John Michel in 1784. However, Michel called these imaginary structures dark stars. In such astronomical bodies, the escape velocity would exceed the speed of light. What began with theoretical stars so dense that not even a single particle of light would be able to escape and as such would not be seen, evolved into the current concept of black holes. The name black hole, however, only emerged in 1963, when John Wheeler, a professor at Princeton, supported by the theory of general relativity, popularized these fabulous physical objects. Since then, black holes have become an important subject to study because black holes allow scientists to study what happens in the most extreme conditions we know of in our universe with regards to temperature, mass, density and time. Thus, studies on black holes have become abundant in scientific literature. In particular, research that deals with the movement of particles closer to black holes is an interesting topic of investigation, as it can reveal the behavior of matter in the presence of intense gravitational fields. One way to describe the movement of particles around hypermassive bodies, probably black holes, is to consider quantum equations in curved spacetime.
 
The analysis of wave propagation opens a perspective on
additional phenomena such as interference effects and radiation scattering in a black hole. Within the scope of quantum mechanics, particles are described by so-called wave functions. In such a way that free particles are effectively described by wave solutions. This subject has been widely discussed in the literature, especially in the case of relativistic particles. In this sense, Damour et al developed a typical work describing the particles in the vicinity of a black hole by solving the Klein-Gordon equation in the Schwrazschild spacetime \cite{damour}, in which the solution of the incident and emitted waves was obtained . The ideas of Damour et al were used by Sannan to construct the particle probability distribution emitted from a black hole \cite{sannan}. In the works cited, Hawking radiation and the evaporation of black holes were treated \cite{hawking_1, hawking_2}. In the context of particles moving at relativistic speeds, other approaches involving Klein-Gordon and Dirac equations in curved spaces are presented in the literature; for instance, the relativistic wave equations are solved for different metrics, such as Kerr, Bertotti-Robinson and AdS black hole \cite{musir, al, cvetic}. In the same aspect, the Klein-Gordon equation is used in the works of Vieira and Bezerra in which resonance frequencies, Hawking radiation and scattering of scalar waves from various types of black holes are investigated, such as the Kerr-Newman-Kasuya (dyon black hole) and Reissner-Nordströn black hole \cite{vieira_1, vieira_2}. The Klein-Gordon equation in curved space can also be used to study the stability of black holes and quasi-normal modes of black holes from various points of view \cite{wheeler, miranda, konoplya_1, konoplya_2, iyer, capistrano, ulhoa} .

There are similar works in the literature involving non-relativistic particles; however, the Schrödinger equation in curved space is generally obtained by the non-relativistic limit of the Dirac or Klein–Gordon equation. This approach allows the generalization of the Schrödinger equation to the curved space \cite{exirifard}. On the other hand, such a generalization leads to a profound question: would it be possible to describe a non-relativistic field through a covariant structure? The answer is positive and the structure that makes this possible is known as Galilean covariance. This approach was introduced by Takahashi based on a convariant version of the Galilei group, which is based on fifth-dimensional tensors \cite{takahashi_1, takahashi_2}. This method was used to analyze non-linearized field equations from which the symmetries of superfluids were discussed in connection with the Goldstone bosons \cite{takahashi_3}. Since then, the notion of Galilean covariance has been developed in several areas, such as cosmology, gravitation and high-spin particle theory in condensed matter \cite{carter_1, carter_2, carter_3, kunzle_1, kunzle_2, ulhoa_2, khanna}. In this work we intend to apply the Galilean covariance method to analyze non-relativistic particles in Schwarzschild-type geometry. Particularly this Schwarzschild-type solution is obtained within the scope of Galilean gravity. That is, Newtonian gravitation is described in a five-dimensional manifold.

The presentation of this work is based on the following topics: in Section 2 we provide a brief review of Galilean covariance; in Section 3 we construct the covariant version of the Schrödinger equation; in Section 4 the Schrödinger equation for Schwarzschild-type spacetime is solved; Section 5 is dedicated to solution analysis; Finally, in Section 6 we present the conclusion and perspectives.

\section{Galilean Covariance}

In this section, we construct the mathematical framework to write the covariant Schrödinger equation. For this propose, we address the galilean covariance. 

In the three dimensional euclidean space $\mathbb{E}_3$ we can define the inner product between two vectors $\mathbf{x}=(x^1,x^2,x^3)$ and $\mathbf{y}=(y^1,y^2,y^3)$ as  
\begin{equation}\label{cov1}
\langle \mathbf{x},\mathbf{y} \rangle=x^1y^1+x^2y^2+x^3y^3.
\end{equation}
Then, the square of norm of vector $\mathbf{x}$ in $\mathbb{E}_3$, denoted by $\|\mathbf{x}\|^2$, is given by
\begin{equation}\label{cov2}
\|\mathbf{x}\|^2=\langle \mathbf{x},\mathbf{x}\rangle=(x^1)^2+(x^2)^2+(x^3)^2.
\end{equation}
Notice that $\|\mathbf{x}\|>0$, i.e., $\|\mathbf{x}\|$ is positive defined. 

In order to construct a nonrelativistic space with dimension up to three we define the inner product between the vectors $x=(x^1,x^2,x^3,x^4,x^5)$ and $y=(y^1,y^2,y^3,y^4,y^5)$ by
\begin{equation}\label{e1}
\langle x, y\rangle= \sum_{i=1}^{3}x^{i}y^{i}-x^4y^5-x^5y^4.
\end{equation}
Eq.(\ref{e1})  can be written by
\begin{equation}\label{e2}
\langle x,y\rangle= \eta_{\mu\nu}x^{\mu}y^{\nu},
\end{equation}
where 
\begin{equation}\label{e3}
\eta_{\mu\nu}=
\begin{pmatrix}
1&0&0&0&0\\
0&1&0&0&0\\
0&0&1&0&0\\
0&0&0&0&-1\\
0&0&0&-1&0
\end{pmatrix}.
\end{equation}
 We notice that $x^{\mu}=\eta^{\mu\nu}x_{\nu}$, $x_{\mu}=\eta_{\mu\nu}x^{\nu}$, and $\eta^{\mu\nu}=(\eta_{\mu\nu})^{-1}$. A vector space $\mathbb{G}_5$ constituted by inner product given in Eq.(\ref{e2}) is a pseudo-riemannian five-dimensional manifold. In this sense, $\eta_{\mu\nu}$ is the metric of $\mathbb{G}_5$. The square of norm of a vector $x$, denoted by $\|x\|^2$, defined in $\mathbb{G}_5$ is given by
 \begin{equation}\label{e4}
 \|x\|^2=\langle x,x\rangle=\|\mathbf{x}\|^2-2x^5x^4.
 \end{equation}
 In the same way in which occurs in the Minkowsky space, we can obtain vectors such as $\|x\|^2>0$,   $\|x\|^2<0$ and  $\|x\|^2=0$. 

 Notice that if we choose $x^5=\frac{\|\mathbf{x}\|^2}{2t}$, $y^5=\frac{\|\mathbf{y}\|^2}{2t}$, $x^4=y^4=t$, Eq.(\ref{e1}) reduces to
 \begin{equation}\label{e5}
 \langle x,y \rangle=\sum_{i=1}^{3}x^{i}y^{i}-\frac{1}{2}\|\mathbf{x}\|^2-\frac{1}{2}\|\mathbf{y}\|^2=-\frac{1}{2}(\|\mathbf{x}\|^2+\|\mathbf{y}\|^2-2\langle \mathbf{x},\mathbf{y}\rangle)=-\frac{1}{2}\langle\mathbf{x-y},\mathbf{x-y}\rangle.
 \end{equation}
 This last result can be considered as an embedding of $\mathbb{E}_3$ in $\mathbb{G}_5$ given by
 \begin{equation}\label{e6}
\mathcal{F}:\mathbf{x}\rightarrow x=\left(\mathbf{x}, x^4, \frac{\|\mathbf{x}\|^2}{2x^4}\right),
 \end{equation}
 where $\mathbf{x} \in \mathbb{E}_3$ and $x \in \mathbb{G}_5$.

 The physical content of this embedding is more clear considering the dispersion relation $\frac{\|\mathbf{p}\|^2}{2m}=E$, i.e., $\|\mathbf{p}\|^2-2mE=0$. In this case, is interesting define the five-momentum vector $p^{\mu}=p=(\mathbf{p},m,E)$, with $\mu=1,..,5$, $\mathbf{p}$ stands the usual three-dimensional momentum, $m$ is the mass, and $E$ is the energy.  Fixing the notation $p^4=E/v$ and $p^5=vm$, in which $v$ is a constant with unity of velocity, then the general five-dimensional dispersion relation can be written as
\begin{equation}\label{e7}
\|\mathbf{p}\|^2-2p^4p^5=\kappa^2,
\end{equation}
where $\kappa$ is a constant. In this work we adopt $v=1$, obtaining $p^4=E$, $p^5=m$, and $\kappa=0$. It is worth noticing that the coordinates of the five-vector $x^{\mu}=x$ defined in section 2 are the 
canonical coordinates associated with the 5-momentum $p^{\mu}$. Then, $\mathbf{x}$ is the canonical 3-vector associated with 3-momentum $\mathbf{p}$, $p^4$ is the canonical coordinate correspondent to $E$, and $p^5$ is the canonical coordinate associated with $m$. In this sense, in accordance with Eq.(\ref{e7}), we define the dispersion relation for coordinates $x^{\mu}x_{\mu}=\eta_{\mu\nu}x^{\mu}x^{\nu}=\|\mathbf{x}\|^2-2x^{4}x^{5}=s^2$. In the special case $p^{\mu}p_{\mu}=0$, we obtain the relations $x^5=\frac{\|\mathbf{x}\|^2}{2t}$, $x^4=t$, i. e., $s=0$. Here it is necessary to clarify that time has no dimension of distance. So a constant with velocity dimension is used to adjust the dimensionality of the coordinate $x^4$, say $v_0$. This constant represents some characteristic speed of the non-relativistic system, such as the speed of sound. It is worth remembering that in the covariant formalism of non-relativistic fields, there is no universal limit speed, such as the speed of light, in principle. But the speed $v_0$ can be identified with $c$ by experimental criteria. Whatever the case, it is possible to adopt a system of units in which $v_0=1$, which we will use unless explicitly stated.

The results presented in this section are useful to construct the covariant Schrödinger equation. 

\section{Covariant Schrödinger Equation}

In this section it is shown how we can obtain the Schrödinger equation from a representation of the Galilei group. From this representation we construct the covariant version for Schrödinger equation.

Galilei group in the covariant notation is defined by the transformations $\mathcal{G}: (x,t)\rightarrow (\overline{x},\overline{t})$ given by
\begin{eqnarray}\nonumber
\overline{x}^i&=&R^{i}_{j}x^{j}+v^{i}x^4+a^i,\\\nonumber
\overline{x}^4&=&x^4+a^4,\\\nonumber
\overline{x}^5&=&x^5+(R^{i}_{j}x^{j})v_i+\frac{1}{2}\mathbf{v}^2x^4,\nonumber
\end{eqnarray}
where $R^{i}_{j}$ stands rotations, $v^{i}$ stands boost, $a^i$ spatial translation, and $a^4$ time translation. The generators of this group can be written by $l_i=\frac{1}{2}\epsilon_{ijk}M_{jk}$, $k_i=M_{5i}$, $c_i=M_{4i}$, $d=M_{54}$, where
$M_{ij}=x_ip_j-x_jp_i$.

Nonrelativistic physical theories, specifically quantum theory, are obtained from unitary and faithful representations of Galilei group. In this context, covariant Schrödinger equation can be obtained from a faithful unitary representation of the Galilei group constructed by the operators defined as ($\hbar=1$)
$\hat{x}_{\mu}=x^\mu$, $\hat{p}_\mu=-i\partial_{\mu}$.
The generators of this group satisfy the non-null commutations relations
\begin{eqnarray}\nonumber
[\hat{l}_i,\hat{l}_j]=i\epsilon_{ijk}l_k, && \qquad [\hat{l}_i,\hat{k}_j]=i\epsilon_{ijk}k_k\\\nonumber
[\hat{l}_i,\hat{c}_j]=i\epsilon_{ijk}c_k, && \quad [\hat{l}_i,\hat{p}_j]=i\epsilon_{ijk}p_k\\\nonumber
[\hat{d},\hat{k}_j]=ik_i,&& \qquad [\hat{c}_i,\hat{d}]=ic_i,\\\nonumber
[\hat{p}_4,\hat{d}]=ip_4,&& \qquad [\hat{d},\hat{p}_5]=ip_5\\\nonumber
[\hat{p}_i,\hat{k}_j]=i\delta_{ij}p_5,&& \qquad [\hat{p}_i,\hat{c}_j]=i\delta_{ij}p_4\\\nonumber
[\hat{p}_4,\hat{k}_i]=ip_i,&& \qquad [\hat{k}_i,\hat{c}_j]=i\epsilon_{ijk}l_k+i\delta_{ij}\hat{d}\\\nonumber
[\hat{p}_5,\hat{c}_i]=ip_i.&&  \\\nonumber
\end{eqnarray}
It is worth noting that the Heisenberg uncertainty relation is obtained from the operators position and momentum, because at this moment we are dealing with a quantum theory.
In this work we consider only spinless representations. In this case, the invariant of the Galilei algebra is given by
\begin{equation}\label{e88}
I_1=\hat{p}^{\mu}\hat{p}_{\mu},
\end{equation}
\begin{equation}\label{e99}
I_2=\hat{p_5}\mathbf{I},
\end{equation}
where $\mathbf{I}$ is the identity operator. Applying these invariant in wave function $\Psi(x)$, we obtain
\begin{equation}\label{e8}
\hat{p}^{\mu}\hat{p}_{\mu}\Psi=k^2\Psi,
\end{equation}
\begin{equation}\label{e9}
\hat{p}_5\Psi=-m\Psi.
\end{equation}
Once $\hat{p}_5=i\partial_5$, we can write $\Psi(x)=\psi(\mathbf{x},x^4)\varphi(x^5)$. Then, the solution of Eq.(\ref{e9}) is given by
\begin{equation}\label{e10}
\varphi(x^5)=e^{imx^5}.
\end{equation}
In this sense, $\Psi(x)=e^{imx^5}\psi(\mathbf{x},x^4)$. Using this last expression in Eq.(\ref{e8}), we obtain
\begin{equation}\label{e11}
i\partial_4\psi(\mathbf{x},x^4)=-\frac{1}{2m}\nabla^2\psi(\mathbf{x},x^4),
\end{equation}
which is the usual Schrödinger equation, since we recognise $x^4=t$. Finally, from Eq.(\ref{e10}) we can write the covariant Schrödinger equation as 
\begin{equation}
-\eta_{\mu\nu}\partial^{\mu}\partial_{\nu}\Psi=k^2\Psi.
\end{equation}
This equation will be generalized to curved space in the next section.

\section{Covariant Schrödinger Equation in Schwarzschild-like\\Spacetime}

The galilean covariance is a formalism that aims to describe non-relativistic fields in a covariant way. For this purpose a five-dimensional manifold is introduced. In the previous section it was shown that the structure of the scalar field in the Galilean manifold reduces to the Schrödinger equation when a dimensional reduction is performed. It should be noted that the curvature of the metric presented in this section vanishes identically. Thus, it is natural to think that non-zero curvature describes a covariant Newtonian gravitation. In fact, a Schwarzschild-type metric is presented in reference \cite{ulhoa_2}, so the scalar field in 5 dimensions can couple with this geometric structure. That is, the result is the description of the Schrödinger equation coupled to Newtonian gravitation. The Galilean Schwarzschild-like line element is given by
\begin{equation}\label{e12}
ds^2=f^{-1}(r)dr^2+r^2d\theta^2+r^2\sin^2\theta d\phi^2-2f(r)dtdx^5,
\end{equation}
where $f(r)=1-2M/r$ and $M$ stands for the non-relativistic black hole mass. In this way the metric tensor can be written as
\begin{equation}\label{e131}
g_{\mu\nu}=
\begin{pmatrix}
f^{-1}(r)&0&0&0&0\\
0&r^2&0&0&0\\
0&0&r^2\sin^2\theta&0&0\\
0&0&0&0&-f(r)\\
0&0&0&-f(r)&0
\end{pmatrix}.
\end{equation}
In this situation, covariant Schrödinger equation can be written as
\begin{equation}\label{e141}
\frac{1}{\sqrt{-g}}\partial_{\mu}(\sqrt{-g}g^{\mu\nu}\partial_{\nu})\Psi=-k^2\Psi,
\end{equation}
with $\sqrt{-g}=r^2\sin\theta$ and $g^{\mu\nu}$ is the inverse tensor of $g_{\mu\nu}$, in which is given by
\begin{equation}\label{e132}
g^{\mu\nu}=
\begin{pmatrix}
f(r)&0&0&0&0\\
0&1/r^2&0&0&0\\
0&0&1/(r^2\sin^2)\theta&0&0\\
0&0&0&0&-f^{-1}(r)\\
0&0&0&-f^{-1}(r)&0
\end{pmatrix}.
\end{equation}

In this way, the following differential equation is obtained
\begin{eqnarray}
  &&\frac{\partial}{\partial r}\left(f(r)\frac{\partial\Psi}{\partial r}\right) + \frac{2}{r}f(r)\frac{\partial \Psi}{\partial r} + \frac{1}{r^2\sin\theta}\frac{\partial}{\partial \theta}\left(\sin\theta \frac{\partial\Psi}{\partial \theta}\right)+ \frac{1}{r^2\sin^2\theta}\frac{\partial^2\Psi}{\partial \phi^2}\nonumber \\
  &&-\frac{1}{f}\frac{\partial^2\Psi}{\partial t\partial x^5}=-k^2\Psi.
\end{eqnarray}
As $\Psi(x)=e^{imx^5}\psi(\mathbf{x},t),$ obtained from equation (\ref{e10}), we get
\begin{eqnarray}
  &&f(r)\left[\frac{\partial}{\partial r}\left(f(r)\frac{\partial\psi}{\partial r}\right) + \frac{2}{r}f(r)\frac{\partial \psi}{\partial r} + \frac{1}{r^2\sin\theta}\frac{\partial}{\partial \theta}\left(\sin\theta \frac{\partial\psi}{\partial \theta}\right)+ \frac{1}{r^2\sin^2\theta}\frac{\partial^2\psi}{\partial \phi^2}\right]\nonumber
  \\
  &&+im\frac{\partial\psi}{\partial t}=-k^2f(r)\psi \label{eq22}
\end{eqnarray}
Now, using the ansatz $\psi(\mathbf{x},t)=e^{-iEt}\Phi(r,\theta, \phi)$, equation~(\ref{eq22}) can be rewritten as
\begin{eqnarray}\nonumber
  &&f(r)\left[\frac{\partial}{\partial r}\left(f(r)\frac{\partial\Phi}{\partial r}\right) + \frac{2}{r}f(r)\frac{\partial \Phi}{\partial r} + \frac{1}{r^2\sin\theta}\frac{\partial}{\partial \theta}\left(\sin\theta \frac{\partial\Phi}{\partial \theta}\right)+ \frac{1}{r^2\sin^2\theta}\frac{\partial^2\Phi}{\partial \phi^2}\right]\\
  &&+(mE+k^2f(r))\Phi=0.\label{eq23}
\end{eqnarray}
Taking \begin{equation}\Phi(r,\theta,\phi)=\sum_{l=0}^{\infty}\sum_{m=-l}^{l}Y_{l,m}(\theta,\phi)\Gamma_{l}(r),\end{equation}
we obtain
\begin{equation}\label{e142}
 \left[\frac{1}{\sin\theta}\frac{\partial}{\partial \theta}\left(\sin\theta \frac{\partial}{\partial \theta}\right)+ \frac{1}{\sin^2\theta}\frac{\partial^2}{\partial \phi^2}\right]Y_{l,m}(\theta,\phi)=-l(l+1)Y_{l,m}(\theta,\phi).
\end{equation}
Simplifying the equation (\ref{eq23}) we get the following equation for the radius $r$:
\begin{equation}
    \left[\frac{mE}{f(r)}+k^2-\frac{l(l+1)}{r^2}+\frac{d}{d r}\left(f(r)\frac{d}{d r}\right) + \frac{2}{r}f(r)\frac{d}{d r}\right]\Gamma_l(r)=0.
    \label{heunc1}
\end{equation}

In order to solve equation (\ref{heunc1}), we write it in the Sturm-Liouville form as:
\begin{equation}
\frac{d}{dr}\left(r(r-2M)\Gamma'_{l}(r)\right)+\left(k^{2}r^{2}-l(l+1)+\frac{mEr^{3}}{r-2M}\right)\Gamma_{l}(r)=0
\label{heunc2}
\end{equation}
Through the transformations $z=1-r/2M$ and \[u(z)=\left[2M(1-z)\right]^{2M\sqrt{-mE}}e^{-2M\sqrt{-mE-k^{2}}(1-z)}\Gamma_l(r)\]
it is possible do rewrite the equation (\ref{heunc2}) as a confluent Heun differential equation~\cite{book:heunc1,book:heunc2,art:plamen} on the form
\[
\frac{d^{2}u}{dz^{2}}+\left(\alpha+\frac{\beta+1}{z}+\frac{\gamma+1}{z-1}\right)\frac{du}{dz}+\left(\frac{\mu}{z}+\frac{\nu}{z-1}\right)u(z)=0,
\]
where,
\begin{eqnarray}
\alpha & =&-4M\sqrt{-mE-k^{2}}\label{const1}\\
\beta & =&-4M\sqrt{-mE}\label{const2}\\
\gamma & =&0\label{const3}\\
\delta & =&-4M^{2}\left(2mE+k^{2}\right)\label{const4}\\
\eta & =&4M^{2}(2mE+k^{2})-l(l+1)\label{const5}\\
\nu & =&\delta-\mu+\alpha\frac{\beta+\gamma+2}{2}\label{const6}\\
\mu & =&\frac{\left(\alpha-\gamma\right)\left(\beta+1\right)-\beta}{2}-\eta\label{const7}
\end{eqnarray}
This transformation gives the solution in terms of confluent Heun functions, valid for \linebreak $0<r<4M$, as:
\begin{eqnarray}
\Gamma_l(r)&=& c_1 \Gamma_{l_f}(r)+c_2\Gamma_{l_\infty}(r) \nonumber \\
&=&c_{1} \left(2 M-r\right)^{2 \sqrt{-m E} M} {\mathrm e}^{\sqrt{-m E-k^{2}} r} \times \nonumber \\
&&\mathtt{HeunC} \left(-4 M \sqrt{-m E-k^{2}},4 \sqrt{-m E} M,0,-4M^2(2 E  m+  k^{2}),\right. \nonumber \\
&&\left. 4M^2(2 E m+4 k^{2})-l(l+1),1-\frac{r}{2 M}\right)\nonumber \\
&+&c_{2} \left(2 M-r\right)^{-2 \sqrt{-m E} M} {\mathrm e}^{\sqrt{-m E-k^{2}}\, r} \times \nonumber \\
&&\mathtt{HeunC} \left(-4 M \sqrt{-m E-k^{2}},-4 \sqrt{-m E}\, M,0,-4M^2(2 E m+ k^{2}),\right. \nonumber \\
&&\left. 4M^2(2 E m+ k^{2})-l(l+1),1-\frac{r}{2 M}\right),
\end{eqnarray}
noting that $\mathtt{HeunC}=\mathtt{HeunC}\left(\alpha,\beta,\gamma,\delta,\eta,w\right)$ is defined for $|w|<1$ and $\alpha, \beta, \gamma, \delta, \eta\in \mathbb{C}$~\cite{book:heunc1,book:heunc2,art:plamen}. Moreover Heun's confluent function is defined satisfying the initial conditions $$\mathtt{HeunC}\left(\alpha,\beta,\gamma,\delta,\eta,0\right)=1$$ and $$\left.\frac{d}{dw}\mathtt{HeunC}\left(\alpha,\beta,\gamma,\delta,\eta,w\right)\right|_{w=0}=\frac{(\gamma-\alpha+1)\beta+\gamma-\alpha+2\eta}{2(\beta+1)}.$$
This result will be used in further sections.


\section{Analysis of Solution}

In this section we will analyze some properties of the solution obtained in the previous section. We will analyze how energy can be quantized and how a particle can leave the event horizon through the tunnel effect. It is worth noting that the solution of the Schrödinger equation coupled to the non-relativistic gravitational field remains restricted to the vicinity of the black hole, as the Heun function is not defined at spatial infinity.

\subsection{Energy Levels near to Schwarzschild-like Radius}
Closer to the non-relativistic Schwarzschild radius,  $r\rightarrow 2M$, the confluent Heun's function approaches to the unity, i.e., $\mathtt{HeunC}\left(\alpha,\beta,\gamma,\delta,\eta,0\right)=1$. Its series expansion for the finite part of the solution, $\Gamma_{l_f}(r)$, gives us:
\begin{eqnarray}
\Gamma_{l_{f}}(r)&=&\left(r-2 M\right)^{2 \sqrt{-2m E}\, M}\times \nonumber \\ &&\left(1-\frac{4M^2(4 E  m+ k^{2})-l(l+1)+2 \sqrt{-2m E}\, M}{2M \left(4 \sqrt{-2m E}\, M+1\right)} \left(r-2 M\right)+O\left((r-2M)^2\right)\right)\quad
\end{eqnarray}
So, for sufficiently small $\varepsilon> 0, $ such that $r-2M=\varepsilon,$ we get the approximation
\begin{equation}\label{energy0}
1-\frac{4M^2(4E  m+ k^{2})-l(l+1)+2 \sqrt{-2m E}\, M}{2M \left(4 \sqrt{-2m E}\, M+1\right)} \varepsilon+O(\varepsilon^2)\cong 1.
\end{equation}
Which gives us
\begin{equation}
\frac{4M^2(4E  m+ k^{2})-l(l+1)+2 \sqrt{-2m E}\, M}{2M \left(4 \sqrt{-2m E}\, M+1\right)}  - O(\varepsilon)\cong 0.
\end{equation}
Then
\begin{equation}\label{energy01}
\frac{4M^2(4E  m+ k^{2})-l(l+1)+2 \sqrt{-2m E}\, M}{2M \left(4 \sqrt{-2m E}\, M+1\right)} \cong 0.
\end{equation}
Finally, solving Eq.\eqref{energy01} for $E$ the following approximation for the energy is obtained:
\begin{equation}\label{energy1}
E_l\cong-\frac{16 M^{2} k^{2}+1\pm\sqrt{32 M^{2} k^{2}-8 l(l+1)+1}-4l(l+1)}{64 M^{2} m}
\end{equation}
for $ l=0,1,2,\ldots$. It should be noted that energy levels $E_l$ depends on quantum number $l$, in addition it is inversely proportional to square of black hole mass.
 
\subsection{Energy Levels for Non-relativistic Super Massives Black Holes}
In this section, we analyze the solution of the Schrödinger equation for Schwarzschild-like spacetime, considering supermassive black holes. In this case, when $r\rightarrow 4M$, It is necessary that the wave the function be finite and well behaved in all points where it is defined. Then, as $r\rightarrow 4M,$ the confluent Heun function must be a polynomial and the following necessary condition will be satisfied in this case~\cite{book:heunc1,book:heunc2,art:plamen}
\begin{equation}\label{energy00}
\frac{\delta}{\alpha}+\frac{\beta}{2}=-n-1, \qquad n=0,1,2,3,\ldots
\end{equation}
Plugging the constant values in equations \eqref{const1}--\eqref{const4} we get
\begin{equation}\label{energy11}
\frac{M(4E m+ k^{2})}{\sqrt{-2m E-k^{2}}}-\frac{4 M\sqrt{-2m E}}{2}=-n-1, \qquad n=0,1,2,3,\ldots
\end{equation}
The condition expressed in Eq.(\ref{energy11}) establishes the quantization of energy of the particle in vicinity of a Schwarzschild-like black hole. In the particular case $k=0$, we obtain
\begin{equation}\label{energy2}
E_n=-\frac{(n+1)^2}{32M^2m}, \qquad n=0,1,2,3,\ldots
\end{equation}
In this sense, we notice that the maximum energy state has energy equal to $E_0=-\frac{1}{32M^2m}$. The negative signal of energy is because the particle of mass $m$ is bounded to the mass $M$. Since $M$ is much larger, those energy levels are very small. It should be noted that energy is defined as being less than a constant. Thus $k$ can be made null without loss of generality.

\subsection{The Tunnel Effect}

In this section, we use the solution given in Eq. (29) to discuss how a particle can pass through the event horizon via quantum tunneling. First, it should be noted that we can consider Eq. (29) as a progressive traveling wave if $-mE<0$, which means a non-bound state. In this case, we have
 \begin{eqnarray}
\Gamma_{l\,\mathrm{in}}(r)&=&\left(2 M-r\right)^{2 i\sqrt{2m E} M} {\mathrm e}^{\sqrt{-2m E} r} \times \nonumber \\
&&\mathtt{HeunC} \left(-4 M \sqrt{-2m E},4 \sqrt{-2m E} M,0,-4M^2(4 E  m),\right. \nonumber\\ 
&&\left. 4M^2(4 E m)-l(l+1),1-\frac{r}{2 M}\right),
\end{eqnarray}
for $r<2M$ and
\begin{eqnarray}
\Gamma_{l\,\mathrm{out}}(r)&=&\left(2 M-r\right)^{-2 i\sqrt{2m E} M} {\mathrm e}^{\sqrt{-2m E} r} \times \nonumber \\
&&\mathtt{HeunC} \left(-4 M \sqrt{-2m E},4 \sqrt{-2m E} M,0,-4M^2(4 E  m),\right. \nonumber \\
&&\left. 4M^2(4 E m)-l(l+1),1-\frac{r}{2 M}\right),
\end{eqnarray}
for $2M<r<4M$. It is here considered $k=0$. The function $\Gamma_{l\,\mathrm{in}}$ represents a progressive wave inside the event horizon, while $\Gamma_{l\,\mathrm{out}}$ is the wave outside the black hole. It should be noted that $\mathtt{HeunC}\left(\alpha,\beta,\gamma,\delta,\eta,0\right)=1$, which is reached at $r\rightarrow 2M$, thus
\begin{eqnarray}\label{haw1}
\Gamma_{l\, \mathrm{in}}(r)&=&\left(2 M-r\right)^{2 i\sqrt{2m E} M} {\mathrm e}^{\sqrt{-2m E} r} \nonumber \\ 
&=&e^{2i\sqrt{2mE}M\left[\log|2M-r|+i\left(2p\pi\right)\right]}e^{\sqrt{-2mE}r}, \qquad p=0,1,2,\ldots
\end{eqnarray}
and 
\begin{eqnarray}
    \Gamma_{l\,\mathrm{out}}(r)&=&\left(2 M-r\right)^{-2 i\sqrt{2m E} M} {\mathrm e}^{\sqrt{-2m E} r} \nonumber \\
&=&e^{-2i\sqrt{2mE}M\left[\log|2M-r|+i\left(\pi+2q\pi\right)\right]}e^{\sqrt{-2mE}r}, \qquad q=0,1,2,\ldots
\end{eqnarray}
Then, there is a transmission coefficient given by
\begin{equation}\label{haw3}
\tau=\frac{|\Gamma_{l\,\mathrm{in}}(r)|^2}{|\Gamma_{l\,\mathrm{out}}(r)|^2}
\end{equation}
which is hence
\begin{eqnarray}\label{haw4}
\tau&=&\left| e^{4i\sqrt{2mE}M [\log|2M-r|+i(\pi+2(p+q)\pi)]} \right|^2 \nonumber \\
&=&e^{-8\pi(2s+1)\sqrt{2mE}M},\qquad s=0,1,2\ldots
\end{eqnarray}
The reflection coefficient can also be written immediately, it reads
\begin{eqnarray}\label{haw5}
N&=&\frac{\tau}{1-\tau}\\\nonumber
&=&\frac{1}{e^{8\pi(2s+1)\sqrt{2mE}M}-1}, \qquad s=0,1,2,\ldots\nonumber
\end{eqnarray}
that is, the coefficient respects the restriction given by a black hole's event horizon, even in non-relativistic physics. However, the transmission coefficient denotes a different prediction from that of black hole physics. Thus a particle described by the progressive wave function can pass through the event horizon  by means of quantum tunneling. It is undeniable that such an effect in relativistic physics would be a mechanism capable of explaining Hawking radiation. On the other hand, in the non-relativistic description there is no such radiation, as the particles do not leave the black hole towards spatial infinity. What occurs, therefore, in the non-relativistic description is an evaporation of the black hole and concentration of matter in the vicinity of the black hole.

\section{Concluding remarks}
In this work we constructed a version for covariant Schrödinger equation for Schwarzschild-like geometry. In sequence, we solved analytically the equation obtained. The solution is given in terms of confluent Heun's function. Using the properties of Heun's function we analyze the solution aimed to study the nonrelativistic particles in vicinity of a Schwarzschild-like black hole. As results we obtained that the energy of particle interacting with black hole is quantized in terms of angular quantum number in the case of ordinary black holes and in terms of a integer number in the case of supermassive black holes. Furthermore, we study the tunnel effect through the black hole's event horizon. We intend to extend the ideas presented in this work to other geometries.


\end{document}